\documentclass[12pt]{article}
\usepackage[centertags]{amsmath}

\begin{document}

\title{\bf Energy-Momentum Problem of Bell-Szekeres Metric in General Relativity
and Teleparallel Gravity}

\author{M. Sharif \thanks{msharif@math.pu.edu.pk} and Kanwal Nazir\\
Department of Mathematics, University of the Punjab,\\
Quaid-e-Azam Campus, Lahore-54590, Pakistan.}

\date{}

\maketitle

\begin{abstract}
This paper is devoted to the investigation of the energy-momentum
problem in two theories, i.e., General Relativity and teleparallel
gravity. We use Einstein, Landau-Lifshitz, Bergmann-Thomson and
M\"{o}ller's prescriptions to evaluate energy-momentum distribution
of Bell-Szekeres metric in both the theories. It is shown that these
prescriptions give the same energy-momentum density components in
both General Relativity and teleparallel theory. M\"{o}ller's
prescription yields constant energy in both the theories.
\end{abstract}


\maketitle

\section{Introduction}

One of the most interesting problems which remains unsolved, since
the birth of General Theory of Relativity (GR), is the
energy-momentum localization. To find a generally accepted
expression, there have been different attempts. However, some
attempts to define the energy-momentum density for the gravitational
field lead to prescriptions that are not true tensors. The first of
such attempts was made by Einstein [1] himself who proposed an
expression for the energy-momentum distribution of the gravitational
field. Following Einstein, many scientists like Landau-Lifshitz [2],
Papapetrou [3], Bergmann-Thomson [4] and M\"{o}ller [5] introduced
their own energy-momentum complexes. All these prescriptions, except
M\"{o}ller, are restricted to do calculations only in Cartesian
coordinates.

The notion of energy-momentum complexes was severely criticized for
a number of reasons. Firstly, the nature of a symmetric and locally
conserved object is non-tensorial and thus its physical
interpretation appeared obscure [6]. Secondly, different
energy-momentum complexes could yield different energy-momentum
distributions for the same gravitational background [7,8]. Finally,
energy-momentum complexes were local objects while it was usually
believed that the suitable energy-momentum of the gravitational
field cannot be localized [9]. An alternate concept of energy,
called quasi-local energy, was developed by Penrose and many others
[10,11]. Although, these quasi-local masses are conceptually very
important, yet these definitions have serious problems. Chang et al.
[12] showed that energy-momentum complexes are actually quasi-local
and legitimate expressions for the energy-momentum.

Virbhadra [13-15] and his collaborator [16] showed that different
energy-momentum complexes yield the same results for a general
non-static spherically symmetric metric of the Kerr-Schild class.
These definitions comply with the quasi-local mass definition of
Penrose for a general non-static spherically symmetric metric of the
Kerr-Schild class. However, these prescriptions disagree in the case
of the most general non-static spherically symmetric metric.
Aguirregabiria et al. [17] proved the consistency of the results
using different energy-momentum complexes for any Kerr-Schild class
metric. Xulu [18,19] extended this investigation and found the same
energy distribution in the Melvin magnetic and Bianchi type I
universes. Chamorro and Virbhadra [20] and Xulu [21] studied the
energy distribution of the charged black holes with a dilaton field.
Ramdinschi and Yang [22], Vagenas [23], Gad [24] and Xulu [25]
investigated the energy distribution of the string black holes using
different prescriptions. Using Einstein and Landau-Lifshitz
complexes, Cooperstock [26] and Rosen [27] found that total energy
of the closed Friedmann-Robertson-Walker (FRW) spacetime vanishes
everywhere. With the help of Einstein energy-momentum complex,
Banerjee and Sen [28] found that the total energy density of Bianchi
type-I universe is zero everywhere. However, there exist examples
[29-32] which do not support this viewpoint.

It has been argued [33,34] that the energy-momentum problem can also
be localized in the framework of the teleparallel theory (TPT) of
gravity. This theory has been considered long time ago in connection
with attempts to define the energy of the gravitational field.
M\"{o}ller [35] was probably the first to notice that the tetrad
description of the gravitational field allows a more satisfactory
treatment of the gravitational energy-momentum than does GR. Using
the teleparallel version of Einstein and Landau-Lifshitz complexes,
Vargas [36] found that total energy of the closed FRW spacetimes
vanishes everywhere. This result agrees with the previous work of
Cooperstock [26] and Rosen [27]. It has been shown [37] that the
results of Bianchi types I and II in TPT are consistent with the
results in GR. Recently, Salti et al. [38,39] have calculated
energy-momentum densities for some particular spacetimes by using
different prescriptions both in GR and TPT and found the same
results. However, there exist some spacetimes that do not provide
consistent results both in GR and TPT. Sharif and Jamil [40]
considered Lewis-Papapetrou metric and found that the results in TPT
do not coincide with those obtained in GR [41].

This paper explores Bell-Szekeres metric by evaluating its
energy-momentum distribution using different prescriptions both in
GR and TPT. We would like to present results rather giving the
details as these are available in the literature [40,41]. The paper
has been organized as follows: Section 2 is devoted to present the
prescriptions of the Einstein, Landau-Lifshitz, Bergmann-Thomson and
M\"{o}ller energy-momentum both in GR and TPT. In section 3, we find
the energy-momentum distribution of Bell-Szekeres metric both in GR
and TPT using these prescriptions. Finally, section 4 provides a
summary and discussion of the results obtained.

\section{Energy-Momentum Complexes}

In this section, we shall briefly outline the Einstein,
Landau-Lifshitz, Bergmann-Thomson and M\"{o}ller prescriptions used
to calculate energy-momentum distribution of a metric both in GR and
TPT.

\subsection{Energy-Momentum Complexes in GR}

For \textbf{Einstein} prescription, the energy-momentum density is
given in the form [5,42]
\begin{equation}
\Theta^{b}_{a}=\frac{1}{16\pi}{H^{bc}_a}_{,c},\quad (a,b,c=0,1,2,3),
\end{equation}
where $H^{bc}_a$= -$H^{cb}_a$ is given by
\begin{equation}
H^{bc}_a=\frac{g_{ad}}{\sqrt{-g}}[-g(g^{bd}g^{ce}-g^{cd}g^{be})]_{,e}.
\end{equation}
Here $g$ is the determinant of the metric tensor $g_{\mu\nu}$ and
comma denotes ordinary differentiation. Notice that $\Theta^0_{0}$
is the energy density, $\Theta^0_i$ $(i=1,2,3)$ are the momentum
density components and $\Theta^i_0$ are the energy current density
components. The momentum four-vector is defined as
\begin{equation}
p_a=\int\int_V \int\Theta^{0}_{a}dx^{1}dx^{2}dx^{3},
\end{equation}
where $p_0$ gives the energy and $p_1$, $p_2$ and $p_3$ are the
momentum components while the integration is taken over the
hypersurface element described by $t=constant$. Einstein's
conservation law becomes
\begin{equation}
\frac{\partial \Theta^{b}_{a}}{\partial x^a}=0.
\end{equation}

\textbf{Landau-Lifshitz} energy-momentum density can be given in
the form [2]
\begin{equation}
L^{ab}=\frac{1}{16\pi}{\ell^{abcd}}_{,cd},
\end{equation}
where
\begin{equation}
\ell^{abcd}=(-g)(g^{ab}g^{cd}-g^{ac}g^{bd}).
\end{equation}
The quantities $L^{00}$ and $L^{0i}$ represent the energy density
and the total momentum (energy current) densities respectively.

The energy-momentum prescription of \textbf{Bergmann-Thomson} is
given by [4]
\begin{equation}
B^{ab}=\frac{1}{16 \pi}{M^{abc}}_{,c},
\end{equation}
where
\begin{equation}
M^{abc}=g^{ad}V^{bc}_d,
\end{equation}
with
\begin{equation}
V^{bc}_d=\frac{g_{de}}{\sqrt{-g}}[-g(g^{be}g^{cf}-g^{ce}g^{bf})]_{,f}.
\end{equation}
The quantities $B^{00}$ and $B^{i0}$ represent energy and momentum
densities respectively. The Bergmann-Thomson energy-momentum
satisfies the following local conservation law
\begin{equation}
\frac{\partial B^{ab}}{\partial x^b}=0
\end{equation}
in any coordinate system. The energy-momentum components are given
by
\begin{equation}
p^a=\int\int_V \int B^{a0}dx^1dx^2dx^3.
\end{equation}
Using Gauss theorem, the above integral takes the form
\begin{equation}
p_a=\frac{1}{16 \pi}\int \int H^{0b}_{a}n_b dS,
\end{equation}
where $n_b$ is the outward unit normal vector to an infinitesimal
surface element $dS$. The quantities $p_i$ give momentum components
while $p_0$ gives the energy.

All the energy-momentum complexes mentioned above are coordinate
dependent and give meaningful results only when the calculations
are carried out in Cartesian coordinates. To overcome this
deficiency, \textbf{M\"{o}ller} [5] introduced another
energy-momentum pseudo-tensor $M^{b}_{a}$ which is coordinate
independent given as
\begin{equation}
M^{b}_{a}=\frac{1}{8\pi}{K^{bc}_{a}}_{,c},
\end{equation}
where
\begin{eqnarray}
K^{bc}_{a}&=&\sqrt{-g}(g_{ad,e}-g_{ae,d})g^{be}g^{cd},
\end{eqnarray}
and $K^{bc}_{a}$ is antisymmetric in its upper indices. This
satisfies the conservation law
\begin{equation}
\frac{\partial M^{b}_{a}}{\partial x^b}=0,
\end{equation}
where $M^0_{0}$ is the energy density, $M^0_i$ are momentum density
components and $M^i_0$ are the components of energy current density.

\subsection{Energy-Momentum Complexes in TPT}

The teleparallel version of Einstein, Bergmann-Thomsan and
Landau-Lifshitz energy-momentum complexes are given [36],
respectively, as
\begin{equation}
hE^{\mu}_{~~\rho}=\frac{1}{4\pi}{U^{~~\mu\nu}_{\rho}}_{,\nu},
\end{equation}
\begin{equation}
hB^{\mu\rho}=\frac{1}{4\pi}[g^{\mu\theta}{U^{~~\rho\nu}_{\theta}}_{,\nu}],
\end{equation}
\begin{equation}
hL^{\mu\rho}=\frac{1}{4\pi}[h g^{\mu\theta}{U^{~~\rho\nu}_{\theta}}_{,\nu}],
\end{equation}
where $h=\det(h^{a}_{~\mu})$ and $U^{~~\mu\nu}_{\rho}$ is the
Freud's superpotential given as
\begin{equation}
U^{~~\mu\nu}_{\rho}=hS^{~~\mu\nu}_{\rho}.
\end{equation}
Here $S^{\rho \mu \nu}$ is the tensor
\begin{eqnarray}
S^{\rho \mu \nu}&=&m_1
T^{\rho\mu\nu}+\frac{m_2}{2}(T^{\mu\rho\nu}-T^{\nu\rho\mu})+
\frac{m_3}{2}(g^{\rho\nu}T^{\theta\mu}_{~~\theta}-g^{\rho\mu}T^{\theta\nu}_{~~\theta}]
\end{eqnarray}
with $m_1,~m_2$ and $m_3$ as the three dimensionless coupling
constants of the teleparallel gravity. For the teleparallel
equivalent of GR, the specific choice of these three constants are
\begin{equation}
m_1=\frac{1}{4},\quad m_2=\frac{1}{2},\quad m_3=-1.
\end{equation}
To calculate this tensor, we evaluate Weitzenb\"{o}ck connection
[43]
\begin{eqnarray}
{\Gamma^\theta}_{\mu\nu}={{h_a}^\theta}\partial_\nu{h^a}_\mu
\end{eqnarray}
which is used to find the corresponding torsion [44]
\begin{equation}
{T^\theta}_{\mu\nu}={\Gamma^\theta}_{\nu\mu}-{\Gamma^\theta}_{\mu\nu}.
\end{equation}
Thus the momentum four-vector for Einstein, Bergmann-Thomsan and
Landau-Lifshitz energy-momentum complexes will be
\begin{equation}
p^{E}_{\mu} =\int_\Sigma h E^{0}_{\mu}dxdydz,
\end{equation}
\begin{equation}
p^{B}_{\mu} =\int_\Sigma h B^{0}_{\mu}dxdydz,
\end{equation}
\begin{equation}
p^{L}_{\mu} =\int_\Sigma h L^{0}_{\mu}dxdydz.
\end{equation}

Now we  discuss M\"{o}ller energy-momentum complex in the context of
TPT. Mikhail et al. [33] defined the superpotential (which is
antisymmetric in its last two indices) of the M\"{o}ller tetrad
theory as
\begin{equation}
U^{\nu\beta}_{\mu}=\frac{\sqrt{-g}}{2\kappa}P^{\tau\nu\beta}_{\chi\rho\sigma}
[\phi^{\rho}g^{\sigma\chi}g_{\mu\tau}-\lambda
g_{\tau\mu}K^{\chi\rho\sigma}-g_{\tau\mu}(1-2\lambda)K^{\sigma\rho\chi}],
\end{equation}
where
\begin{equation}
P^{\tau\nu\beta}_{\chi\rho\sigma}
=\delta^{\tau}_{\chi}g^{\nu\beta}_{\rho\sigma}
+\delta^{\tau}_{\rho}g^{\nu\beta}_{\sigma\chi}
-\delta^{\tau}_{\sigma}g^{\nu\beta}_{\chi\rho},
\end{equation}
and $g^{\nu\beta}_{\rho\sigma}$ is a tensor quantity defined by
\begin{equation}
g^{\nu\beta}_{\rho\sigma}=\delta^{\nu}_{\rho}\delta^{\beta}_{\sigma}
-\delta^{\nu}_{\sigma}\delta^{\beta}_{\rho}.
\end{equation}
Here $K^{\sigma\rho\chi}$ is a contortion tensor, $\lambda$ is a
free dimensionless coupling constant of TPT, $\kappa$ is the
coupling constant and $\phi_{\mu}$ is the basis vector field given
by
\begin{equation}
\phi_{\mu}=T^{\nu}_{~~\nu\mu}.
\end{equation}
The energy-momentum density is defined as
\begin{equation}
M^{\nu}_{\mu}={U^{~~\nu\rho}_{\mu}}_{,\rho}.
\end{equation}
The energy $E$ contained in a sphere of radius $R$ is expressed by
the volume integral as
\begin{equation}
p_{\mu}(R)=\int_{r=R} \int\int {U^{0\rho}_{\mu}}_{,\rho}dx^{3},
\end{equation}
\begin{equation}
p_{\mu}(R)=\int_{r=R} \int\int M^{0}_{\mu}dx^{3}.
\end{equation}

\section{Bell-Szekeres Metric}

It is well-known that exact plane gravitational waves are simple
time dependent plane symmetric solutions of the Einstein field
equations [45]. The colliding plane wave spacetimes have been
investigated extensively in GR [46] due to their interesting
behaviour. The first exact solution of the Einstein-Maxwell
equations representing colliding plane shock electromagnetic waves
with co-linear polarizations was obtained by Bell and Szekeres [47].
This solution is conformally flat in the interaction region and is
represented by the metric
\begin{equation}
ds^{2}= 2du dv+e^{-U}(e^V dx^2+e^{-V}dy^2),
\end{equation}
where the metric functions $U$ and $V$ depend on the null
coordinates $u$ and $v$. The complete solution of the
Einstein-Maxwell equations is
\begin{eqnarray}
U&=&-\log(f(u)+g(u)), \nonumber\\
V&=&\log(r w-p q)-\log(r w+p q),
\end{eqnarray}
where
\begin{eqnarray}
r&=&(\frac{1}{2}+f)^\frac{1}{2},\quad p=(\frac{1}{2}-f)^\frac{1}{2}, \nonumber\\
w&=&(\frac{1}{2}+g)^\frac{1}{2},\quad
q=(\frac{1}{2}-g)^\frac{1}{2}
\end{eqnarray}
with
\begin{eqnarray}
f&=&\frac{1}{2}-\sin^2P,\quad g=\frac{1}{2}-\sin^2Q.
\end{eqnarray}
Here $P=au\theta(u),~Q=bv\theta(v)$, where $\theta$ is the Heaviside
unit step function, $a$ and $b$ are arbitrary constants.

The Cartesian form of the metric is found by substituting $t=u+v$
and $z=v-u$
\begin{eqnarray}
ds^{2}&=&\frac{1}{2}dt^{2}-\cos^2\{\frac{1}{2} b(t-z)
\theta(\frac{t-z}{2})+\frac{1}{2} a(t+z)
\theta(\frac{t+z}{2})\}dx^2\nonumber\\&-&\cos^2\{\frac{1}{2} b(z-t)
\theta(\frac{t-z}{2})+\frac{1}{2} a(t+z) \theta(\frac{t+z}{2})\}dy^2
-\frac{1}{2}dz^{2}.\nonumber\\
\end{eqnarray}

\subsection{Energy-Momentum Distribution in GR}

In this section, we find the energy-momentum distribution of
Bell-Szekeres metric using Einstein, Landau-Lifshitz,
Bergmann-Thomson and M\"{o}ller's prescriptions in GR.

Using \textbf{Einstein} prescription, the components of the
energy-momentum density of Bell-Szekeres metric turn out to be
\begin{eqnarray}
\Theta^{00}&=&\frac{1}{32
\pi}[\cos\{b(t-z)\theta(\frac{t-z}{2})\}\{2b
\theta(\frac{t-z}{2})+b(t-z) \theta'(\frac{t-z}{2})\}^2\nonumber\\
&+&b \sin\{b(t-z)
\theta(\frac{t-z}{2})\}\{4\theta'(\frac{t-z}{2})+(t-z)
\theta''(\frac{t-z}{2})\}\nonumber\\
&+&a\{a \cos\{a(t+z)\theta(\frac{t+z}{2})\}\{2\theta(\frac{t+z}{2})\nonumber\\
&+&(t+z)\theta'(\frac{t-z}{2})\}^2+\sin\{a(t+z)\theta(\frac{t+z}{2})\}\{4
\theta'(\frac{t+z}{2})\nonumber\\
&+&(t+z)\theta''(\frac{t+z}{2})\}\}],\nonumber\\
\Theta^{10}&=&\Theta^{20}=0,\nonumber\\
\Theta^{30}&=&\frac{1}{32
\pi}[\cos\{b(t-z)\theta(\frac{t-z}{2})\}\{2b
\theta(\frac{t-z}{2})+b(t-z)
\theta'(\frac{t-z}{2})\}^2\nonumber\\
&+&b
\sin\{b(t-z)\theta(\frac{t-z}{2})\}\{4\theta'(\frac{t-z}{2})+(t-z)
\theta''(\frac{t-z}{2})\}\nonumber\\
&+&a\{-a \cos\{a(t+z)\theta(\frac{t+z}{2})\}\{2 \theta(\frac{t+z}{2})\nonumber\\
&+&(t+z)\{\theta'(\frac{t-z}{2})\}^2-\sin\{a(t+z)
\theta(\frac{t+z}{2})\}\{4
\theta'(\frac{t+z}{2})\nonumber\\
&+&(t+z)\theta''(\frac{t+z}{2})\}\}\}].
\end{eqnarray}

When we use \textbf{Landau-Lifshitz} complex, we obtain the
following energy-momentum density components
\begin{eqnarray}
L^{00}&=&\frac{1}{256\pi}[2\cos\{2b(t-z)\theta(\frac{t-z}
{2})\}\{2b\theta(\frac{t-z}{2})
+b(t-z)\theta'(\frac{t-z}{2})\}^2\nonumber\\
&+&b\sin\{2b(t-z)\theta(\frac{t-z}{2})\}
\{4\theta'(\frac{t-z}{2})+(t-z)\theta''(\frac{t
-z}{2})\}\nonumber\\
&+&2b
\sin\{b(t-z)\theta(\frac{t-z}{2})\}\{2a\sin(a(t+z)
\theta(\frac{t+z}{2}))(2\theta(\frac{t
-z}{2})\nonumber\\
&+&(t-z)\theta'(\frac{t-z}{2}))(2\theta(\frac{t-z}{2})
+(t+z)\theta'(\frac{t-z}{2}))\nonumber
\end{eqnarray}
\begin{eqnarray}
&+&\cos(a(t+z)\theta(\frac{t+z}{2}))(4\theta'
(\frac{t-z}{2})+(t-z)\theta''(\frac{t-z}{2}))\}\nonumber\\
&+&2\cos\{b(t-z)\theta(\frac{t-z}{2})\}\{cos\{a(t+z)\theta(\frac{t+z}{2})\{4a^2
\theta^2(\frac{t+z}{2})\nonumber\\
&+&\{2b\theta(\frac{t-z}{2})+b(t-z)\theta'
(\frac{t-z}{2})\}^2+4a^2(t+z)\theta(\frac{t
+z}{2})\theta'(\frac{t+z}{2})\nonumber\\
&+&a^2(t+z)^2\theta'(\frac{t+z}{2})^2\}+a
\sin\{a(t+z)\theta(\frac{t+z}{2})\}\{4
\theta'(\frac{t+z}{2})\nonumber\\
&+&(t+z)\theta''(\frac{t+z}{2})\}\}\}+a\{2a
\cos \{2a(t+z)\theta(\frac{t+z}{2})\}
\{2\theta(\frac{t+z}{2})\nonumber\\
&+&(t+z)\theta'(\frac{t+z}{2})\}^2
+\sin\{2a(t+z)\theta(\frac{t+z}{2})\}
\{4\theta'(\frac{t+z}{2})\nonumber\\
&+&(t+z)\theta''(\frac{t+z}{2})\}\}],\nonumber\\
L^{10}&=&L^{20}=0,\nonumber\\
L^{30}&=&\frac{1}{256
\pi}[2\cos\{2b(t-z)\theta(\frac{t-z}{2})\}\{2b\theta(\frac{t-z}{2})
+b(t-z)\theta'(\frac{t-z}{2})\}^2\nonumber\\
&+&2b\cos\{a(t+z)\theta(\frac{t+z}{2})\}
\sin\{b(t-z)\theta(\frac{t-z}{2}))\}\{4\theta'(\frac{t
-z}{2})\nonumber\\
&+&(t-z)\theta''(\frac{t-z}{2})\}+b
\sin\{2b(t-z)\theta(\frac{t-z}{2})\}\{4\theta'(\frac{t-z}{2})\nonumber\\
&+&(t-z)\theta''(\frac{t-z}{2})\}+2
\cos\{b(t-z)\theta(\frac{t-z}{2})\}\cos\{a(t\nonumber\\
&+&z)\theta(\frac{t+z}{2})\}\{2b\theta(\frac{t-z}{2})-2a
\theta(\frac{t+z}{2})+b(t-z)\theta'( \frac{t-z}{2})\nonumber\\
&-&a(t+z)\theta'(\frac{t+z}{2})\}\{2b \theta(\frac{t-z}{2})+2a
\theta(\frac{t+z}{2})
+b(t-z)\theta'(\frac{t-z}{2})\nonumber\\
&+&a(t+z)\theta'(\frac{t+z}{2})\}-a \sin\{a(t+z)
\theta(\frac{t+z}{2})\}\{4 \theta'(\frac{t+z}{2})\nonumber\\
&+&(t+z)\theta''(\frac{t+z}{2})\}+a\{-2a
\cos\{2a(t+z)\theta(\frac{t+z}{2})\}\{2\theta(\frac{t+z}{2})\nonumber\\
&+&(t+z)\theta'(\frac{t+z}{2})\}^2-
\sin\{2a(t+z)\theta(\frac{t+z}{2})\}\{4
\theta'(\frac{t+z}{2})\nonumber\\
&+&(t+z)\theta''(\frac{t+z}{2})\}\}].
\end{eqnarray}

The energy-momentum distribution in \textbf{Bergmann-Thomson's}
prescription will become
\begin{eqnarray}
B^{00}&=&\frac{1}{32 \pi}[\cos\{b(t-z)\theta(\frac{t-z}{2})\}\{2b
\theta(\frac{t-z}{2})+b(t-z) \theta'(\frac{t-z}{2})\}^2\nonumber
\end{eqnarray}
\begin{eqnarray}
&+&b \sin\{b(t-z)
\theta(\frac{t-z}{2})\}\{4\theta'(\frac{t-z}{2})+(t-z)
\theta''(\frac{t-z}{2})\}\nonumber\\
&+&a\{a \cos\{a(t+z)\theta(\frac{t+z}{2})\}\{2
\theta(\frac{t+z}{2})+(t+z)\theta'(\frac{t-z}{2})\}^2\nonumber\\
&+&\sin\{a(t+z)\theta(\frac{t+z}{2})\}\{4 \theta'(\frac{t+z}{2})
+(t+z)\theta''(\frac{t+z}{2})\}\}],\nonumber\\
B^{10}&=&B^{20}=0,\nonumber\\
B^{30}&=&\frac{1}{32 \pi}[ \cos\{b(t-z)\theta(\frac{t-z}{2})\}\{2b
\theta(\frac{t-z}{2})+b(t-z)
\theta'(\frac{t-z}{2})\}^2\nonumber\\
&+&b
\sin\{b(t-z)\theta(\frac{t-z}{2})\}\{4\theta'(\frac{t-z}{2})+(t-z)
\theta''(\frac{t-z}{2})\}\nonumber\\
&+&a\{-a \cos\{a(t+z)\theta(\frac{t+z}{2})\}\{2
\theta(\frac{t+z}{2})+(t+z)\{\theta'(\frac{t-z}{2})\}^2\nonumber\\
&-&\sin\{a(t+z) \theta(\frac{t+z}{2})\}\{4 \theta'(\frac{t+z}{2})\nonumber\\
&+&(t+z)\theta''(\frac{t+z}{2})\}\}\}].
\end{eqnarray}

Finally, energy and momentum densities in \textbf{M\"{o}ller's}
prescription take the form
\begin{eqnarray}
M^{a0}=0,\quad(a=0,1,2,3)
\end{eqnarray}

\subsection{Energy-Momentum Distribution in TPT}

The tetrad components, in Cartesian coordinates, of Eq.(38) are
given as
\begin{eqnarray}
{h^{0}}_{0}&=& \frac{1}{\sqrt{2}},\nonumber\\
{h^{1}}_{1}&=&\cos\{\frac{1}{2}b(t-z)\theta(\frac{t-z}{2})
+\frac{1}{2}a(t+z)\theta(\frac{t+z}{2})\},\nonumber\\
{h^{2}}_{2}&=& \cos\{\frac{1}{2}b(t-z)\theta(\frac{t-z}{2})
-\frac{1}{2}a(t+z)\theta(\frac{t+z}{2})\},\nonumber\\
{h^{3}}_{3}&=& \frac{1}{\sqrt{2}}
\end{eqnarray}
and its inverse is
\begin{eqnarray}
{h_{0}}^{0}&=& \sqrt{2} ,\nonumber\\
{h_{1}}^{1}&=& \sec[\frac{1}{2}\{b(t-z)\theta(\frac{t-z}{2})
+a(t+z)\theta(\frac{t+z}{2})\}],\nonumber
\end{eqnarray}
\begin{eqnarray}
{h_{2}}^{2}&=& \sec[\frac{1}{2}\{b(t-z)\theta(\frac{t-z}{2})
-a(t+z)\theta(\frac{t+z}{2})\}],\nonumber\\
{h_{3}}^{3}&=& \sqrt{2}.
\end{eqnarray}
Using these values, we can find the Weitzenb$\ddot{o}$ck connections
and the corresponding torsion tensor. These are then used to find
the components of the superpotential that are essential to obtain
energy-momentum density components. When we make use of the
components of superpotential in Eqs.(16)-(18), the energy-momentum
density components of \textbf{Einstein, Landau-Lifshitz and
Bergmann-Thomson} become
\begin{eqnarray}
hE^{00}&=&\frac{1}{32 \pi}[\cos\{b(t-z)\theta(\frac{t-z}{2})\}\{2b
\theta(\frac{t-z}{2})+b(t-z) \theta'(\frac{t-z}{2})\}^2\nonumber\\
&+&b \sin\{b(t-z)
\theta(\frac{t-z}{2})\}\{4\theta'(\frac{t-z}{2})+(t-z)
\theta''(\frac{t-z}{2})\}\nonumber\\
 &+&a\{a \cos\{a(t+z)\theta(\frac{t+z}{2})\}\{2
\theta(\frac{t+z}{2})\nonumber\\
&+&(t+z)\theta'(\frac{t-z}{2})\}^2+\sin\{a(t+z)\theta(\frac{t+z}{2})\}\{4
\theta'(\frac{t+z}{2})\nonumber\\
&+&(t+z)\theta''(\frac{t+z}{2})\}\}],\nonumber\\
hE^{10}&=&hE^{20}=0,\nonumber\\
hE^{30}&=&\frac{1}{32 \pi}[\cos\{b(t-z)\theta(\frac{t-z}{2})\}\{2b
\theta(\frac{t-z}{2})+b(t-z) \theta'(\frac{t-z}{2})\}^2\nonumber\\
&+&b\sin\{b(t-z)\theta(\frac{t-z}{2})\}\{4\theta'(\frac{t-z}{2})+(t-z)
\theta''(\frac{t-z}{2})\}\nonumber\\
&+&a\{-a \cos\{a(t+z)\theta(\frac{t+z}{2})\}\{2
\theta(\frac{t+z}{2})\nonumber\\
&+&(t+z)\{\theta'(\frac{t-z}{2})\}^2-\sin\{a(t+z)
\theta(\frac{t+z}{2})\}\{4
\theta'(\frac{t+z}{2})\nonumber\\
&+&(t+z)\theta''(\frac{t+z}{2})\}\}\}],\nonumber\\
hL^{00}&=&\frac{1}{256\pi}[2
\cos\{2b(t-z)\theta(\frac{t-z}{2})\}\{2b\theta(\frac{t-z}{2})
+b(t-z)\theta'(\frac{t-z}{2})\}^2\nonumber\\
&+&b
\sin\{2b(t-z)\theta(\frac{t-z}{2})\}\{4\theta'(\frac{t-z}{2})+(t-z)
\theta''(\frac{t-z}{2})\}\nonumber\\
&+&2b \sin\{b(t-z)\theta(\frac{t-z}{2})\}\{2a
\sin\{a(t+z)\theta(\frac{t+z}{2})\}\{2\theta(\frac{t-z}{2})\nonumber\\
&+&(t-z)\theta'(\frac{t-z}{2})\}\{2\theta(\frac{t-z}{2})
+(t+z)\theta'(\frac{t-z}{2})\}\nonumber
\end{eqnarray}
\begin{eqnarray}
&+& \cos\{a(t+z)\theta(\frac{t+z}{2})\}\{4\theta'(\frac{t-z}{2})
+(t-z)\theta''(\frac{t-z}{2})\}\}\nonumber\\
&+&2
\cos\{b(t-z)\theta(\frac{t-z}{2})\}\{\cos\{a(t+z)\theta(\frac{t+z}{2})\{4a^2
\theta^2(\frac{t+z}{2})\nonumber\\
&+&\{2b\theta(\frac{t-z}{2})+b(t-z)\theta'(\frac{t-z}{2})\}^2+4a^2(t+z)\theta(\frac{t
+z}{2})\theta'(\frac{t+z}{2})\nonumber\\
&+&a^2(t+z)^2\theta'(\frac{t+z}{2})^2\}+a
\sin\{a(t+z)\theta(\frac{t+z}{2})\{4
\theta'(\frac{t+z}{2})\nonumber\\
&+&(t+z)\theta''(\frac{t+z}{2})\}\}+a\{2a
\cos\{2a(t+z)\theta(\frac{t+z}{2})\}
\{2\theta(\frac{t+z}{2})\nonumber\\
&+&(t+z)\theta'(\frac{t+z}{2})\}^2+\sin\{2a(t+z)
\theta(\frac{t+z}{2})\{4\theta'(\frac{t+z}{2})\nonumber\\
&+&(t+z)\theta''(\frac{t+z}{2})\}\}\}\}\}],\nonumber\\
hL^{10}&=&hL^{20}=0,\nonumber\\
hL^{30}&=&\frac{1}{256 \pi}[2
\cos\{2b(t-z)\theta(\frac{t-z}{2})\}\{2b\theta(\frac{t-z}{2})
+b(t-z)\theta'(\frac{t-z}{2})\}^2\nonumber\\
&+&2b\cos\{a(t+z)\theta(\frac{t+z}{2})\}
\sin\{b(t-z)\theta(\frac{t-z}{2})\}\{4\theta'
(\frac{t-z}{2})\nonumber\\
&+&(t-z)
\theta''(\frac{t-z}{2})\}+b\sin\{2b(t-z)
\theta(\frac{t-z}{2})\}\{4\theta'(\frac{t-z}{2})
+(t-z)\theta''(\frac{t-z}{2})\}\nonumber\\
&+&2\cos\{b(t-z)\theta(\frac{t-z}{2})\}\cos\{a(t+z)\theta
(\frac{t+z}{2})\}\{2b\theta(\frac{t-z}{2})-2a\theta(\frac{t+z}{2})\nonumber\\
&+&b(t-z)\theta'(\frac{t-z}{2})-a(t+z)\theta'(\frac{t+z}{2})\}
(2b \theta(\frac{t-z}{2})+2a \theta(\frac{t+z}{2})\nonumber\\
&+&b(t-z)\theta'(\frac{t-z}{2})+a(t+z)\theta'(\frac{t+z}{2}))-a
\sin\{a(t+z) \theta(\frac{t+z}{2})\}\{4
\theta'(\frac{t+z}{2})\nonumber\\
&+&(t+z)\theta''(\frac{t+z}{2})\}+a\{-2a
\cos\{2a(t+z)\theta(\frac{t+z}{2})\}\{2\theta(\frac{t+z}{2})\nonumber\\
&+&(t+z)\theta'(\frac{t+z}{2})\}^2-
\sin\{2a(t+z)\theta(\frac{t+z}{2})\}\{4
\theta'(\frac{t+z}{2})\nonumber\\
&+&(t+z)\theta''(\frac{t+z}{2})\}\}],\nonumber
\end{eqnarray}
and
\begin{eqnarray}
hB^{00}&=&\frac{1}{32 \pi}[\cos\{b(t-z)\theta(\frac{t-z}{2})\}\{2b
\theta(\frac{t-z}{2})+b(t-z) \theta'(\frac{t-z}{2})\}^2\nonumber\\
&+&b \sin\{b(t-z)
\theta(\frac{t-z}{2})\}\{4\theta'(\frac{t-z}{2})+(t-z)
\theta''(\frac{t-z}{2})\}\nonumber
\end{eqnarray}
\begin{eqnarray}
&+&a\{a \cos(a(t+z)\theta(\frac{t+z}{2}))(2
\theta(\frac{t+z}{2})\nonumber\\
&+&(t+z)\theta'(\frac{t-z}{2}))^2+
\sin(a(t+z)\theta(\frac{t+z}{2}))\{4
\theta'(\frac{t+z}{2})\nonumber\\
&+&(t+z)\theta''(\frac{t+z}{2})\}\}],\nonumber\\
hB^{10}&=&hB^{20}=0,\nonumber\\
hB^{30}&=&\frac{1}{32 \pi}[\cos\{b(t-z)\theta(\frac{t-z}{2})\}\{2b
\theta(\frac{t-z}{2})+b(t-z)
\theta'(\frac{t-z}{2})\}^2\nonumber\\
&+&b
\sin\{b(t-z)\theta(\frac{t-z}{2})\}\{4\theta'(\frac{t-z}{2})+(t-z)
\theta''(\frac{t-z}{2})\}\nonumber\\
&+&a\{-a
\cos\{a(t+z)\theta(\frac{t+z}{2})\}\{2 \theta(\frac{t+z}{2})\nonumber\\
&+&(t+z)(\theta'(\frac{t-z}{2}))^2-\sin\{a(t+z)\theta(\frac{t+z}{2})\}\{4
\theta'(\frac{t+z}{2})\nonumber\\
&+&(t+z)\theta''(\frac{t+z}{2})\}\}\}],
\end{eqnarray}
respectively.

Similarly, we can proceed to find energy-momentum density components
using \textbf{M\"{o}ller's} prescription in TPT. The tetrad
components of Eq.(34) are given as
\begin{eqnarray}
{h^a}_\mu=
\begin{pmatrix}
1 & \frac{1}{2} & 0 & 0 \\
-1 & \frac{1}{2} & 0 & 0 \\
0 & 0 & \cos\{au \theta(u)+ b v \theta(v)\} & 0 \\
0 & 0 & 0 & \cos\{au \theta(u)- b v \theta(v)\}
\end{pmatrix}
\end{eqnarray}
and its inverse
\begin{eqnarray}
{h_a}^\mu=
\begin{pmatrix}
\frac{1}{2} & 1 & 0 & 0 \\
-\frac{1}{2} & 1 & 0 & 0 \\
0 & 0 & \sec\{au \theta(u)+ b v \theta(v)\} & 0 \\
0 & 0 & 0 & \sec\{au \theta(u)- b v \theta(v)\}
\end{pmatrix}.
\end{eqnarray}
Consequently, the energy-momentum density components turn out to be
\begin{eqnarray}
M^{00}&=&M^{20}=M^{30}=0,\nonumber\\
M^{10}&=&\frac{a}{\kappa}[2a \cos\{2a u
\theta(u)\}\{\theta(u)+u \theta'(u)\}^2\nonumber\\
&+& \sin\{2a u \theta(u)\}\{2 \theta'(u)+u \theta''(u)\}].
\end{eqnarray}

\section{Discussion}

The problem of energy-momentum localization has been a subject of
many researchers but still remains un-resolved. Numerous attempts
have been made to explore a quantity which describes the
distribution of energy-momentum due to matter, non-gravitational
and gravitational fields. This paper continues the investigation
of comparing various distributions presented in the literature in
the framework of GR and TPT. The more information that is
assembled on the subject, the better. We have used four different
prescriptions namely Einstein, Landau-Lifshitz, Bergmann and
M\"{o}ller to calculate energy-momentum distribution of a
Bell-Szekeres metric in the context of both GR and TPT. The
resulting non-vanishing components of the energy-momentum density
are displayed in the following tables. In these tables, EMD will
stand for energy-momentum density.

\vspace{0.5cm}

\noindent {\bf {\small Table 1(a)}} {\bf Bell-Szekeres Metric:
Einstein's Prescription in GR and TPT}

\vspace{0.1cm}

\begin{center}
\begin{tabular}{|l|l|}
\hline {\bf EMD} & {\bf Expression}
\\ \hline $\Theta^{00}$ & $
\begin{array}{c}
\frac{1}{32
\pi}[\cos\{b(t-z)\theta(\frac{t-z}{2})\}\{2b\theta(\frac{t-z}{2})
+b(t-z) \theta'(\frac{t-z}{2})\}^2\\
+b \sin\{b(t-z) \theta(\frac{t-z}{2})\}\{4\theta'(\frac{t-z}{2})
+(t-z)\theta''(\frac{t-z}{2})\}\\
+a\{a \cos\{a(t+z)\theta(\frac{t+z}{2})\}\{2\theta(\frac{t+z}{2})
+(t+z)\theta'(\frac{t-z}{2})\}^2\\
+\sin\{a(t+z)\theta(\frac{t+z}{2})\}\{4 \theta'(\frac{t+z}{2})
+(t+z)\theta''(\frac{t+z}{2})\}\}]
\end{array}$
\\ \hline $\Theta^{30}$ & $\begin{array}{c}
\frac{1}{32 \pi}[\cos\{b(t-z)\theta(\frac{t-z}{2})\}\{2b
\theta(\frac{t-z}{2})
+b(t-z)\theta'(\frac{t-z}{2})\}^2\\
+b \sin\{b(t-z)\theta(\frac{t-z}{2})\}\{4\theta'(\frac{t-z}{2})
+(t-z)\theta''(\frac{t-z}{2})\}\\
+a\{-a \cos\{a(t+z)\theta(\frac{t+z}{2})\}\{2
\theta(\frac{t+z}{2})
+(t+z)\{\theta'(\frac{t-z}{2})\}^2\\
-\sin\{a(t+z) \theta(\frac{t+z}{2})\}\{4 \theta'(\frac{t+z}{2})
+(t+z)\theta''(\frac{t+z}{2})\}\}\}]
\end{array}$
\\ \hline
\end{tabular}
\end{center}

\vspace{0.3cm}

\begin{center}
\begin{tabular}{|l|l|}
\hline {\bf EMD} & {\bf Expression}
\\ \hline $hE^{00}$ & $
\begin{array}{c}
\frac{1}{32
\pi}[\cos\{b(t-z)\theta(\frac{t-z}{2})\}\{2b\theta(\frac{t-z}{2})
+b(t-z) \theta'(\frac{t-z}{2})\}^2\\
+b \sin\{b(t-z) \theta(\frac{t-z}{2})\}\{4\theta'(\frac{t-z}{2})
+(t-z)\theta''(\frac{t-z}{2})\}\\
+a\{a \cos\{a(t+z)\theta(\frac{t+z}{2})\}\{2\theta(\frac{t+z}{2})
+(t+z)\theta'(\frac{t-z}{2})\}^2\\
+\sin\{a(t+z)\theta(\frac{t+z}{2})\}\{4 \theta'(\frac{t+z}{2})
+(t+z)\theta''(\frac{t+z}{2})\}\}]
\end{array}$
\\ \hline $hE^{30}$ & $\begin{array}{c}
\frac{1}{32 \pi}[\cos\{b(t-z)\theta(\frac{t-z}{2})\}\{2b
\theta(\frac{t-z}{2})
+b(t-z)\theta'(\frac{t-z}{2})\}^2\\
+b \sin\{b(t-z)\theta(\frac{t-z}{2})\}\{4\theta'(\frac{t-z}{2})
+(t-z)\theta''(\frac{t-z}{2})\}\\
+a\{-a \cos\{a(t+z)\theta(\frac{t+z}{2})\}\{2
\theta(\frac{t+z}{2})
+(t+z)\{\theta'(\frac{t-z}{2})\}^2\\
-\sin\{a(t+z) \theta(\frac{t+z}{2})\}\{4 \theta'(\frac{t+z}{2})
+(t+z)\theta''(\frac{t+z}{2})\}\}\}]
\end{array}$
\\ \hline
\end{tabular}
\end{center}
\newpage
\vspace{0.5cm}

\noindent {\bf {\small Table 1(b)}} {\bf Bell-Szekeres Metric:
Landau-Lifshitz's Prescription in GR and TPT}

\vspace{0.9cm}

\begin{tabular}{|l|l|}
\hline {\bf EMD} & {\bf Expression}
\\ \hline $L^{00}$ & $
\begin{array}{c}
\frac{1}{256\pi}[2\cos\{2b(t-z)\theta(\frac{t-z}{2})\}\{2b\theta(\frac{t-z}{2})
+b(t-z)\theta'(\frac{t-z}{2})\}^2\\
+b\sin\{2b(t-z)\theta(\frac{t-z}{2})\}\{4\theta'(\frac{t-z}{2})
+(t-z)\theta''(\frac{t-z}{2})\}\\
+2b \sin\{b(t-z)\theta(\frac{t-z}{2})\}\{2a\sin(a(t+z)
\theta(\frac{t+z}{2}))(2\theta(\frac{t-z}{2})\\
+(t-z)\theta'(\frac{t-z}{2}))(2\theta(\frac{t-z}{2})
+(t+z)\theta'(\frac{t-z}{2}))
+\cos(a(t\\
+z)\theta(\frac{t+z}{2}))(4\theta'(\frac{t-z}{2})
+(t-z)\theta''(\frac{t-z}{2}))\}
+2\cos\{b(t\\
-z)\theta(\frac{t-z}{2})\}\{cos\{a(t+z)\theta(\frac{t+z}{2})\{4a^2
\theta^2(\frac{t+z}{2})
+\{2b\theta(\frac{t-z}{2})\\
+b(t-z)\theta'(\frac{t-z}{2})\}^2
+4a^2(t+z)\theta(\frac{t+z}{2})\theta'(\frac{t+z}{2})
+a^2(t+z)^2\\
\theta'(\frac{t+z}{2})^2\}+a
\sin\{a(t+z)\theta(\frac{t+z}{2})\}\{4\theta'(\frac{t+z}{2})\\
+(t+z)\theta''(\frac{t+z}{2})\}\}\}\\
+a\{2a\cos
\{2a(t+z)\theta(\frac{t+z}{2})\}\{2\theta(\frac{t+z}{2})
+(t+z)\theta'(\frac{t+z}{2})\}^2\\
+\sin\{2a(t+z)\theta(\frac{t+z}{2})\}\{4\theta'(\frac{t +z}{2})
+(t+z)\theta''(\frac{t+z}{2})\}\}]
\end{array}$
\\ \hline $L^{30}$ & $
\begin{array}{c}
\frac{1}{256
\pi}[2\cos\{2b(t-z)\theta(\frac{t-z}{2})\}\{2b\theta(\frac{t-z}{2})
+b(t-z)\theta'(\frac{t-z}{2})\}^2\\
+2b\cos\{a(t+z)\theta(\frac{t+z}{2})\}\sin\{b(t-z)\theta(\frac{t-z}{2}))\}
\{4\theta'(\frac{t-z}{2})\\
+(t-z)\theta''(\frac{t-z}{2})\}\\
+b\sin\{2b(t-z)\theta(\frac{t-z}{2})\}\{4\theta'(\frac{t-z}{2})+(t-z)
\theta''(\frac{t-z}{2})\}\\
+2\cos\{b(t-z)\theta(\frac{t-z}{2})\}\cos\{a(t+z)\theta(\frac{t+z}{2})\}\\
\{2b\theta'(\frac{t-z}{2})-2a
\theta(\frac{t+z}{2})+b(t-z)\theta'(\frac{t-z}{2})
-a(t+z)\theta'(\frac{t+z}{2})\}\\
\{2b \theta(\frac{t-z}{2})+2a \theta(\frac{t+z}{2})
+b(t-z)\theta'(\frac{t-z}{2})+a(t+z)\theta'(\frac{t+z}{2})\}\\
-a \sin\{a(t+z)\theta(\frac{t+z}{2})\}\{4\theta'(\frac{t+z}{2})
+(t+z)\theta''(\frac{t+z}{2})\}\\
+a\{-2a\cos\{2a(t+z)\theta(\frac{t+z}{2})\}\{2\theta(\frac{t+z}{2})
+(t+z)\theta'(\frac{t+z}{2})\}^2\\-
\sin\{2a(t+z)\theta(\frac{t+z}{2})\}\{4 \theta'(\frac{t+z}{2})
+(t+z)\theta''(\frac{t+z}{2})\}\}]
\end{array}$
\\ \hline
\end{tabular}

\vspace{0.1cm}
\begin{tabular}{|l|l|}
\hline {\bf EMD} & {\bf Expression}
\\ \hline $hL^{00}$ & $
\begin{array}{c}
\frac{1}{256\pi}[2\cos\{2b(t-z)\theta(\frac{t-z}{2})\}\{2b\theta(\frac{t-z}{2})
+b(t-z)\theta'(\frac{t-z}{2})\}^2\\
+b\sin\{2b(t-z)\theta(\frac{t-z}{2})\}\{4\theta'(\frac{t-z}{2})
+(t-z)\theta''(\frac{t-z}{2})\}\\
+2b \sin\{b(t-z)\theta(\frac{t-z}{2})\}\{2a\sin(a(t+z)
\theta(\frac{t+z}{2}))(2\theta(\frac{t-z}{2})\\
+(t-z)\theta'(\frac{t-z}{2}))(2\theta(\frac{t-z}{2})
+(t+z)\theta'(\frac{t-z}{2}))+\cos(a(t\\
+z)\theta(\frac{t+z}{2}))(4\theta'(\frac{t-z}{2})
+(t-z)\theta''(\frac{t-z}{2}))\}+2\cos\{b(t\\
-z)\theta(\frac{t-z}{2})\}\{cos\{a(t+z)\theta(\frac{t+z}{2})\{4a^2
\theta^2(\frac{t+z}{2})
+\{2b\theta(\frac{t-z}{2})\\
+b(t-z)\theta'(\frac{t-z}{2})\}^2
+4a^2(t+z)\theta(\frac{t+z}{2})\theta'(\frac{t+z}{2})
+a^2(t+z)^2\\
\theta'(\frac{t+z}{2})^2\}+a
\sin\{a(t+z)\theta(\frac{t+z}{2})\}\{4\theta'(\frac{t+z}{2})\\
+(t+z)\theta''(\frac{t+z}{2})\}\}\}\\
+a\{2a\cos
\{2a(t+z)\theta(\frac{t+z}{2})\}\{2\theta(\frac{t+z}{2})
+(t+z)\theta'(\frac{t+z}{2})\}^2\\
+\sin\{2a(t+z)\theta(\frac{t+z}{2})\}\{4\theta'(\frac{t +z}{2})
+(t+z)\theta''(\frac{t+z}{2})\}\}]
\end{array}$
\\ \hline $hL^{30}$ & $
\begin{array}{c}
\frac{1}{256
\pi}[2\cos\{2b(t-z)\theta(\frac{t-z}{2})\}\{2b\theta(\frac{t-z}{2})
+b(t-z)\theta'(\frac{t-z}{2})\}^2\\
+2b\cos\{a(t+z)\theta(\frac{t+z}{2})\}\sin\{b(t-z)\theta(\frac{t-z}{2}))\}
\{4\theta'(\frac{t-z}{2})\\
+(t-z)\theta''(\frac{t-z}{2})\}\\
+b\sin\{2b(t-z)\theta(\frac{t-z}{2})\}\{4\theta'(\frac{t-z}{2})+(t-z)
\theta''(\frac{t-z}{2})\}\\
+2\cos\{b(t-z)\theta(\frac{t-z}{2})\}\cos\{a(t+z)\theta(\frac{t+z}{2})\}\\
\{2b\theta'(\frac{t-z}{2})-2a
\theta(\frac{t+-z}{2})+b(t-z)\theta'(\frac{t-z}{2})
-a(t+z)\theta'(\frac{t+z}{2})\}\\
\{2b \theta(\frac{t-z}{2})+2a \theta(\frac{t+z}{2})
+b(t-z)\theta'(\frac{t-z}{2})+a(t+z)\theta'(\frac{t+z}{2})\}\\
-a \sin\{a(t+z)\theta(\frac{t+z}{2})\}\{4\theta'(\frac{t+z}{2})
+(t+z)\theta''(\frac{t+z}{2})\}\\
+a\{-2a\cos\{2a(t+z)\theta(\frac{t+z}{2})\}\{2\theta(\frac{t+z}{2})
+(t+z)\theta'(\frac{t+z}{2})\}^2\\-
\sin\{2a(t+z)\theta(\frac{t+z}{2})\}\{4 \theta'(\frac{t+z}{2})
+(t+z)\theta''(\frac{t+z}{2})\}\}]
\end{array}$
\\ \hline
\end{tabular}

\vspace{0.5cm}

\noindent {\bf {\small Table 1(c)}} {\bf Bell-Szekeres Metric:
Bergmann's Prescription in GR and TPT}

\vspace{0.1cm}

\begin{center}
\begin{tabular}{|l|l|}
\hline {\bf EMD} & {\bf Expression}
\\ \hline $B^{00}$ & $
\begin{array}{c}
\frac{1}{32
\pi}[\cos\{b(t-z)\theta(\frac{t-z}{2})\}\{2b\theta(\frac{t-z}{2})
+b(t-z) \theta'(\frac{t-z}{2})\}^2\\
+b \sin\{b(t-z) \theta(\frac{t-z}{2})\}\{4\theta'(\frac{t-z}{2})
+(t-z)\theta''(\frac{t-z}{2})\}\\
+a\{a \cos\{a(t+z)\theta(\frac{t+z}{2})\}\{2\theta(\frac{t+z}{2})
+(t+z)\theta'(\frac{t-z}{2})\}^2\\
+\sin\{a(t+z)\theta(\frac{t+z}{2})\}\{4 \theta'(\frac{t+z}{2})
+(t+z)\theta''(\frac{t+z}{2})\}\}]
\end{array}$
\\ \hline $B^{30}$ & $\begin{array}{c}
\frac{1}{32 \pi}[\cos\{b(t-z)\theta(\frac{t-z}{2})\}\{2b
\theta(\frac{t-z}{2})
+b(t-z)\theta'(\frac{t-z}{2})\}^2\\
+b \sin\{b(t-z)\theta(\frac{t-z}{2})\}\{4\theta'(\frac{t-z}{2})
+(t-z)\theta''(\frac{t-z}{2})\}\\
+a\{-a \cos\{a(t+z)\theta(\frac{t+z}{2})\}\{2
\theta(\frac{t+z}{2})
+(t+z)\{\theta'(\frac{t-z}{2})\}^2\\
-\sin\{a(t+z) \theta(\frac{t+z}{2})\}\{4 \theta'(\frac{t+z}{2})
+(t+z)\theta''(\frac{t+z}{2})\}\}\}]
\end{array}$
\\ \hline
\end{tabular}
\end{center}

\begin{center}
\begin{tabular}{|l|l|}
\hline {\bf EMD} & {\bf Expression}
\\ \hline $hB^{00}$ & $
\begin{array}{c}
\frac{1}{32
\pi}[\cos\{b(t-z)\theta(\frac{t-z}{2})\}\{2b\theta(\frac{t-z}{2})
+b(t-z) \theta'(\frac{t-z}{2})\}^2\\
+b \sin\{b(t-z) \theta(\frac{t-z}{2})\}\{4\theta'(\frac{t-z}{2})
+(t-z)\theta''(\frac{t-z}{2})\}\\
+a\{a \cos\{a(t+z)\theta(\frac{t+z}{2})\}\{2\theta(\frac{t+z}{2})
+(t+z)\theta'(\frac{t-z}{2})\}^2\\
+\sin\{a(t+z)\theta(\frac{t+z}{2})\}\{4 \theta'(\frac{t+z}{2})
+(t+z)\theta''(\frac{t+z}{2})\}\}]
\end{array}$
\\ \hline $hB^{30}$ & $\begin{array}{c}
\frac{1}{32 \pi}[\cos\{b(t-z)\theta(\frac{t-z}{2})\}\{2b
\theta(\frac{t-z}{2})
+b(t-z)\theta'(\frac{t-z}{2})\}^2\\
+b \sin\{b(t-z)\theta(\frac{t-z}{2})\}\{4\theta'(\frac{t-z}{2})
+(t-z)\theta''(\frac{t-z}{2})\}\\
+a\{-a \cos\{a(t+z)\theta(\frac{t+z}{2})\}\{2
\theta(\frac{t+z}{2})
+(t+z)\{\theta'(\frac{t-z}{2})\}^2\\
-\sin\{a(t+z) \theta(\frac{t+z}{2})\}\{4 \theta'(\frac{t+z}{2})
+(t+z)\theta''(\frac{t+z}{2})\}\}\}]
\end{array}$
\\ \hline
\end{tabular}
\end{center}

\vspace{0.5cm}

\noindent {\bf {\small Table 1(d)}} {\bf Bell-Szekeres Metric:
M\"{o}ller's Prescription in GR and in TPT}

\vspace{0.5cm}

Energy and momentum become constant for M\"{o}ller's prescription in
GR.

\vspace{0.5cm}

\begin{center}
\begin{tabular}{|l|l|}
\hline {\bf EMD} & {\bf Expression}
\\ \hline $M^{10}$ & $
\begin{array}{c}
\frac{a}{\kappa}[2a \cos\{2a u \theta(u)\}\{\theta(u)+u \theta'(u)\}^2\\
+ \sin\{2a u \theta(u)\}\{2 \theta'(u)+u \theta''(u)\}]
\end{array}$
\\ \hline
\end{tabular}
\end{center}
Notice that energy is constant in both GR and TPT.

From these tables, it follows that the energy-momentum density
components turn out to be finite and well-defined in each case. It
is interesting to note that we obtain same results for all the four
prescriptions used here both in GR and TPT. We find constant energy
for M\"{o}ller's prescription in both these theories. The summary of
the results is the following:
\begin{eqnarray*}
E^{a0}&=&hE^{a0}=B^{a0}=hB^{a0},\nonumber\\
L^{a0}&=&hL^{a0},\nonumber\\
M^0_0&=&hM^0_0,\quad (a=0,1,2,3).
\end{eqnarray*}
This indicates that both GR and TPT are equivalent theories for
Einstein, Bergmann-Thomson, Landau-Lifshitz and M\"{o}ller's
prescriptions.

It has been shown [48] that for a given spacetime many quasi-local
mass definitions do not give agreed results. In GR, several
energy-momentum expressions (reference frame dependent
pseudo-tensors) have been proposed. All this makes it difficult to
decide which one to use, and raises a suspicion that they could give
different energy-momentum distributions for one fixed spacetime. We
can conclude that the use of the energy-momentum complexes may not
be sufficient to find the energy-momentum distribution of the
physical systems.

Finally, we would like to mention here that the tetrad formalism
itself has some advantages. These advantages come mainly from the
independence of the tetrad formalism from the equivalence principle
and consequent suitability to the discussion of quantum issues. Some
classic solutions of the field equations have already been
translated into the teleparallel language. Thus TPT seems to provide
a more appropriate environment to deal with the energy problem.

\vspace{2cm}

{\bf \large References}

\begin{description}

\item{[1]} Trautman, A.: Gravitation: {\it An Introduction to
Current Research} ed. Witten, L. (Wiley, New York, 1962)169.

\item{[2]} Landau, L.D. and Lifshitz, E.M.: {\it The Classical
Theory of Fields} (Addison-Wesley Press, New York, 1962).

\item{[3]} Papapetrou, A.: {\it Proc. R. Irish Acad}.
\textbf{A52}(1948)11.

\item{[4]} Bergmann, P.G. and Thompson, R.: Phys. Rev.
\textbf{D89}(1958)400.

\item{[5]} M\"{o}ller, C.: Ann. Phys. (NY) \textbf{4}(1958)347.

\item{[6]} Chandrasekhar, S. and Ferrari, V.: {\it Proc. Roy. Soc. London}
\textbf{A435}(1991)645.

\item{[7]} Bergqvist, G.: Class. Quantum Gravit. \textbf{9}(1992)1753.

\item{[8]} Bergqvist, G.: Class. Quantum Gravit. \textbf{9}(1992)1917.

\item{[9]} Nester, J.M. and Chen, C.: Class. Quantum Gravit. \textbf{16}(1999)1279.

\item{[10]} Penrose, R.: {\it Proc. Roy. Soc. London}
\textbf{A388}(1982)457; GR 10 \textit{Conference} eds. Bertotti,
B., de Felice, F. and Pascolini, A. I(Padova, 1983)607.

\item{[11]} Brown, J.D. and York, Jr. J.W.: Phys. Rev. \textbf{
D47}(1993)1407.

\item{[12]} Chang, C.C., Nester, J.M. and Chen, C.: Phys. Rev.
Lett. \textbf{83}(1999)1897.

\item{[13]}  Virbhadra, K.S.: Phys. Rev. \textbf{ D41}(1990)1086;
\textbf{D42}(1990)1066 and references therein.

\item{[14]} Virbhadra, K.S.: Phys. Rev. \textbf{ D60}(1999)104041.

\item{[15]} Virbhadra, K.S.: Phys. Rev. \textbf{ D42}(1990)2919.

\item{[16]} Rosen, N. and Virbhadra, K.S.: Gen. Relativ. Gravit.
\textbf{25}(1993)429.

\item{[17]} Aguirregabiria, J.M., Chamorro, A. and Virbhadra, K.S.:
Gen. Relativ. Gravit. \textbf{28}(1996)1393.

\item{[18]} Xulu, S.S.: Int. J. Mod. Phys.
\textbf{A15}(2000)2979; Mod. Phys. Lett. \textbf{A15}(2000)1151 and
reference therein.

\item{[19} Xulu, S.S.: Astrophys. Space Sci. \textbf{283}(2003)23.

\item{[20]} Chamorro, A. and Virbhardra, K.S.: Int. J. Mod.
Phys. \textbf{D5}(1994)251.

\item{[21]} Xulu, S.S.: Int. J. Mod. Phys. \textbf{D7}(1998)773.

\item{[22]} Ramdinschi, I. and Yang, I.C.: \textit{ On the Energy
of String Black Holes, New Developments in String Theory Research }
ed. Grece, A. (New York, Nova Science, 2005).

\item{[23]} Vagenas, E.C.: Int. J. Mod. Phys.
\textbf{A18}(2003)5781.

\item{[24]} Gad, R.M.: Astrophys. Space Sci. \textbf{295}(2005)459.

\item{[25]} Xulu, S.S.: Int. J. Theor. Phys. \textbf{37}(2003)1773.

\item{[26]} Cooperstock, F.I.: Mod. Phys. Lett.
\textbf{A14}(1999)1531.

\item{[27]} Rosen, N.: Gen. Relativ. Gravit. \textbf{26}(1994)323.

\item{[28]} Banerjee, N and Sen, S.: Pramana J. Phys.
\textbf{49}(1997)609.

\item{[29]} Sharif, M.: Int. J. Mod. Phys.
\textbf{A17}(2002)1175.

\item{[30]} Sharif, M.: Int. J. Mod. Phys.
\textbf{A18}(2003)4361.

\item{[31]} Sharif, M. and Fatima, T.: Int. J. Mod. Phys.
\textbf{A20}(2005)4309.

\item{[32]} Sharif, M. and Fatima, T.: Nouvo Cim.
\textbf{B120}(2005)533.

\item{[33]} Mikhail, F.I., Wanas, M.I., Hindawi, A. and Lashin,
E.I.: Int. J. Theo. Phys. \textbf{32}(1993)1627.

\item{[34]} Nashed, G.G.L.: Nouvo Cim. \textbf{B117}(2002)521.

\item{[35]} M\"{o}ller, C.: {\it Tetrad Fields and Conservation
Laws in General Relativity} (Academic Press, London, 1962).

\item{[36]} Vargas, T.: Gen. Relativ. Gravit. \textbf{30}(2004)1255.

\item{[37]} Loi So, L. and Vargas, T.: Chinese J. Phys.
\textbf{43}(2005)901.

\item{[38]} Salti, M. and Havare, A.: Int. J. Mod. Phys.
\textbf{A20}(2005)2169.

\item{[39]} Aydogdu, O. and Salti, M.: Astrophys. Space Sci.
\textbf{229}(2005)227.

\item{[40]} Sharif, M. and Amir, M.J.: Mod. Phys. Lett. \textbf{A22}(2007)425..

\item{[41]} Sharif, M. and Azam, M.: Int. J. Mod. Phys.
\textbf{A22}(2007)1935.

\item{[42]} M\"{o}ller, C.: Ann. Phys. (NY) \textbf{12}(1961)118.

\item{[43]} Aldrovendi, R. and Pereira, J.G.: {\it An Introduction to
Gravitation Theory} (preprint)

\item{[44]} Aldrovandi and Pereira, J.G.: {\it An Introduction to
Geometrical Physics} (World Scientific, 1995).

\item{[45]} Baldwin, O.R. and Jeffrey, G.B.: \textit{Proc. R. Soc. London}
\textbf{A111}(1926)95.

\item{[46]} Griffiths, J.B.: \textit{Colliding Plane Waves in General Relativity}
(Clarendon Press, Oxford, 1991).

\item{[47]} Bell, P. and Szekeres, P.: Gen. Relativ. Gravit. \textbf{5}(1974)275.

\item{[48]} Bergqvist, G.: Class. Quantum Gravit.
\textbf{9}(1992)1753 .

\end{description}

\end{document}